\documentstyle[12pt]{article}

\newcommand{\sect}[1]{\setcounter{equation}{0}\section{#1}}

\def\bseq{\begin{subequation}}  % = 1a 1b
\def\eseq{\end{subequation}}
\def\bsea{\begin{subeqnarray}}  % = 1.1a 1.1b
\def\esea{\end{subeqnarray}}

\def\Dot#1{{\kern0.5pt
     {#1} \kern-5.05pt \raise5.8pt\hbox{$\textstyle.$}\kern
0.5pt}}

\newcommand{\beq}{\begin{equation}}
\newcommand{\eeq}{\end{equation}}
\newcommand{\bea}{\begin{eqnarray}}
\newcommand{\eea}{\end{eqnarray}}
\newcommand{\ena}{\end{eqnarray}}

\newcommand {\non}{\nonumber}

\renewcommand{\a}{\alpha}
\renewcommand{\b}{\beta}

\renewcommand{\d}{\delta}
\newcommand{\pa}{\partial}

\newcommand{\g}{\gamma}
\newcommand{\G}{\Gamma}

\newcommand{\e}{\epsilon}

\renewcommand{\l}{\lambda}

\newcommand{\m}{\mu}

\newcommand{\s}{\sigma}

\renewcommand{\o}{\omega}

\def\Mb{\kern 2pt\mathchoice
	    {%displaystyle
	     \vbox{\hrule width10pt height 0.4pt depth 0pt
		 \kern 1.2pt\hbox{\kern -2pt$\displaystyle M$}}}
	    {%textstyle
		 \vbox{\hrule width10pt height 0.4pt depth 0pt
		 \kern 1.2pt\hbox{\kern -2pt$\textstyle M$}}}
	    {%scriptstyle \kern 0.5pt
\vbox{\hrule width6pt height 0.4pt depth 0pt
		 \kern 1.0pt\hbox{\kern -2pt$\scriptstyle M$}}}
	    {%scriptscriptstyle \kern 0.5pt
		 \vbox{\hrule width5pt height 0.4pt depth 0pt
		 \kern 0.8pt\hbox{\kern -2pt$\scriptscriptstyle M$}}}}

\def\Sb{\kern 2pt\mathchoice
	    {%displaystyle
		 \vbox{\hrule width6pt height 0.4pt depth 0pt
		 \kern 1.2pt\hbox{\kern -2pt$\displaystyle S$}}}
	    {%textstyle
		 \vbox{\hrule width6pt height 0.4pt depth 0pt
		 \kern 1.2pt\hbox{\kern -2pt$\textstyle S$}}}
	    {%scriptstyle
		 \vbox{\hrule width3.5pt height 0.4pt depth 0pt
		 \kern 1.0pt\hbox{\kern -2pt$\scriptstyle S$}}}
	    {%scriptscriptstyle
		 \vbox{\hrule width3pt height 0.4pt depth 0pt
		 \kern 0.8pt\hbox{\kern -2pt$\scriptscriptstyle S$}}}}

\def\Rb{\kern 2pt\mathchoice
	    {%displaystyle
		 \vbox{\hrule width5.5pt height 0.4pt depth 0pt
		 \kern 1.2pt\hbox{\kern -2.5pt$\displaystyle R$}}}
	    {%textstyle
		 \vbox{\hrule width5.5pt height 0.4pt depth 0pt
		 \kern 1.2pt\hbox{\kern -2.5pt$\textstyle R$}}}
	    {%scriptstyle
		 \vbox{\hrule width3.5pt height 0.4pt depth 0pt
		 \kern 1.0pt\hbox{\kern -2.2pt$\scriptstyle R$}}}
	    {%scriptscriptstyle
		 \vbox{\hrule width3pt height 0.4pt depth 0pt
		 \kern 0.8pt\hbox{\kern -2.2pt$\scriptscriptstyle R$}}}}

  \def\pp{{\mathchoice
	  {%displaystyle
	      \kern 1pt%
	      \raise 1pt
	      \vbox{\hrule width5pt height0.4pt depth0pt
		    \kern -2pt
		    \hbox{\kern 2.3pt
			  \vrule width0.4pt height6pt depth0pt
			  }
		    \kern -2pt
		    \hrule width5pt height0.4pt depth0pt}%
		    \kern 1pt
	   }
	    {%textstyle
	      \kern 1pt%
	      \raise 1pt
	      \vbox{\hrule width4.3pt height0.4pt depth0pt
		    \kern -1.8pt
		    \hbox{\kern 1.95pt
			  \vrule width0.4pt height5.4pt depth0pt
			  }
		    \kern -1.8pt
		    \hrule width4.3pt height0.4pt depth0pt}%
		    \kern 1pt
	    }
	    {%scriptstyle
	      \kern 0.5pt%
	      \raise 1pt
	      \vbox{\hrule width4.0pt height0.3pt depth0pt
		    \kern -1.9pt  %[e]=0.15pt
		    \hbox{\kern 1.85pt
			  \vrule width0.3pt height5.7pt depth0pt
			  }
		    \kern -1.9pt
		    \hrule width4.0pt height0.3pt depth0pt}%
		    \kern 0.5pt
	    }
	    {%scriptscriptstyle
	      \kern 0.5pt%
	      \raise 1pt
	      \vbox{\hrule width3.6pt height0.3pt depth0pt
		    \kern -1.5pt
		    \hbox{\kern 1.65pt
			  \vrule width0.3pt height4.5pt depth0pt
			  }
		    \kern -1.5pt
		    \hrule width3.6pt height0.3pt depth0pt}%
		    \kern 0.5pt%}
	    }
	}}

  \def\mm{{\mathchoice
		       {%displaystyle
			     \kern 1pt
	       \raise 1pt    \vbox{\hrule width5pt height0.4pt depth0pt
				  \kern 2pt
				  \hrule width5pt height0.4pt depth0pt}
			     \kern 1pt}
		       {%textstyle
			    \kern 1pt
	       \raise 1pt \vbox{\hrule width4.3pt height0.4pt depth0pt
				  \kern 1.8pt
				  \hrule width4.3pt height0.4pt depth0pt}
			     \kern 1pt}
		       {%scriptstyle
			    \kern 0.5pt
	       \raise 1pt
			    \vbox{\hrule width4.0pt height0.3pt depth0pt
				  \kern 1.9pt
				  \hrule width4.0pt height0.3pt depth0pt}
			    \kern 1pt}
		       {%scriptscriptstyle
			   \kern 0.5pt
	     \raise 1pt  \vbox{\hrule width3.6pt height0.3pt depth0pt
				  \kern 1.5pt
				  \hrule width3.6pt height0.3pt depth0pt}
			   \kern 0.5pt}
		       }}

\def\pd{{\kern0.5pt
		   + \kern-5.05pt \raise5.8pt\hbox{$\textstyle.$}\kern 
0.5pt}}

\def\pmd{{\kern0.5pt
		  \pm \kern-5.05pt
\raise6.3pt\hbox{$\textstyle.$}\kern1.5pt}}

\def\md{{\mathchoice
   {%displaystyle
      {{\kern 1pt - \kern-6.2pt \raise5pt\hbox{$\textstyle.$}\kern
1pt}}}
    {%textstyle
      {{\kern 1pt - \kern-6.2pt \raise5pt\hbox{$\textstyle.$}\kern
1pt}}}
    {%scriptstyle
      {\kern0.5pt - \kern-5.05pt
\raise3.4pt\hbox{$\textstyle.$}\kern0.5pt}}
    {%scriptscriptstyle
      {\kern0.5pt - \kern-5.05pt
\raise3.4pt\hbox{$\textstyle.$}\kern0.5pt}}}}

\newcommand{\Del}{\nabla}

 \newcommand{\tr}{{\rm tr}}

\begin{document}

\begin{titlepage}
{\hbox to\hsize{November 2000 \hfill
{BRX TH-482}}}
{\hbox to\hsize{${~}$ \hfill
{McGill 00-30}}}
\begin{center}
\vglue .04in
{\Large \bf Norcor vs the Abominable Gauge Completion}
\footnote{Supported in 
part by National Science Foundation Grant 
PHY-00-70475.}
\\[.15in]
Marcus T. Grisaru \footnote{grisaru@brandeis.edu}\\
{\it Physics Department, Brandeis University\\
Waltham, MA 02454 USA}
\\[.1in]
Marcia E. Knutt\footnote{knutt@physics.mcgill.ca}\\
{\it Physics Department, McGill University \\
Montreal, QC CANADA H3A 2T8}
\\[1in]

{\bf ABSTRACT}\\[.0015in]
\end{center}

We use normal coordinate methods to obtain the expansion
with respect to fermionic coordinates of the 11-dimensional
supermembrane action in a supergravity background.  Likewise,
expansions  for various branes in other
dimensions can be obtained. These methods allow  a
systematic and  unambiguous  expansion of the vielbein to any order in the
fermionic coordinates and avoid the complications encountered in the gauge completion approach.

${~~~}$ \newline
PACS: 04.65.+e,11.90.+t,12.60.Jv \\[.01in]  
Keywords: Superspace, Supergravity, Membranes.
\end{titlepage}

\section{Introduction}

~~~~In the past few years, in the course of investigations of
properties of branes in supergravity backgrounds, several authors
\cite{Bernard, Bernardtalk, Nicolai, Plefka, Wati, Cvetic, Peeters} have performed
expansions of supermembrane actions with respect to the
fermionic coordinates $\theta^\a$. For such expansions, which
involve in particular  expanding the vielbein $E^A(x, \theta)$, they
have employed  an abomination known  as ``gauge completion'' ({\it pace} friends),
a procedure which  manages to cast component results in a superspace
language by laboriously comparing order-by-order  component supersymmetry 
transformations with superspace coordinate transformations. This
procedure is complicated, ambiguous, unmanageable at higher
orders, and just plain ugly. It may be unavoidable
 in some cases where the appropriate superspace formalism
does not exist or is not sufficiently developed. (We note that for some specific backgrounds
 compact  explicit form of the vielbein can be obtained. This is the case for   AdS$_5\times$S$^5$
    \cite{Kallosh}, as well as AdS$_4\times$S$^7$  and AdS$_7\times$S$^4$ \cite{dewit, Bernardtalk}; see also ref. \cite{Siegel}.)
 However, in all cases where a geometric description is available in terms of
superspace covariant derivatives with torsions and curvatures satisfying suitable constraints
 a much more elegant
and efficient method exists, based on normal coordinate expansions
of the vielbein and all other quantities \cite{ Macarthur, Dhar, sigmamodel, bfunction,
norcor1, norcor2}. This method has been used primarily for the expansion
of the Green-Schwarz superstring action  \cite{Dhar, sigmamodel, bfunction},
or for the derivation of the superspace density formula \cite{norcor1, norcor2},
but it can be readily applied to the expansion of various brane actions.
In this paper we demonstrate its use for the case of the 11-dimensional
supermembrane. In a later publication we shall consider various
branes in 10-dimensional supergravity backgrounds.

In the following sections we first summarize the basic elements of
11-dimensional supergravity and the description of the  supermembrane.
We then give the general form of the normal coordinate expansion and specialize to the
present case, writing down the expansion of the vielbein components ${E_M}^A$ through the
first three orders in the fermionic coordinates. We identify the  component fields (graviton, gravitino and three-form field
strength, as well as their (component) covariant derivatives) that appear in the expansion, and finally give the  low-order expansion of
the membrane, specializing eventually to a bosonic background.

\section{11-dimensional supergravity}
~~~~The theory is described in a superspace with coordinates $Z^M = (x^m, \theta^\mu)$ by the vielbein ${E}^A (x, \theta) = dZ^M {E_M}^A$
and three-form $B =(1/3!)E^CE^BE^A B_{ABC}$  satisfying torsion constraints
 and field-strength constraints 
respectively, as follows \cite{Howe, Cremmer}:

\newpage
\bea
(a)~~~&&{T_{\a \b}}^c = -i (\Gamma^c)_{\a \b} \nonumber\\
(b)~~~&&{T_{\a \b}}^{\gamma} = {T_{\a b}}^c ={T_{a b}}^c=0\nonumber\\
(c)~~~&&H_{\a \b \gamma \d} = H_{\a \b \gamma d}=H_{\a bcd}=0 \nonumber\\
(d)~~~&&H_{\a \b cd}= i (\Gamma_{cd})_{\a \b}
\label{constraints}
\eea
with $H = dB =(1/4!)E^DE^CE^BE^AH_{ABCD}$ and
\beq
H_{ABCD}= \sum_{(ABCD)}\Del_A B_{BCD} +{T_{AB}}^E B_{ECD}
\eeq
 (We use real Majorana
gamma--matrices $\Gamma^a$, and $\Gamma^{abc...}$ antisymmetrized with unit
strength.) These constraints put the theory on shell.

{}From the Bianchi identities $DT^A = E^B{R_B}^A$, $D{R_A}^B=0$ and
$dH=0$, or 
\bea
&&\sum_{(ABC)}({R_{ABC}}^D- \Del_A{T_{BC}}^D-{T_{AB}}^E {T_{EC}}^D)=0
\nonumber\\
&&\sum_{(ABCD)}(\Del_A{R_{BCD}}^E+{T_{AB}}^F{R_{FCD}}^E)=0
\\
&&\sum_{(ABCDE)}(\Del_AH_{BCDE} +{T_{AB}}^F H_{FCDE})=0
 %-\sum_{(ADBEC)}(-1)^{BC +BD +CE}{T_{AD}}^F H_{FBEC}
\nonumber
\label{Bianchis}
\eea
one derives expressions for the remaining components of the torsion
and the  components of the curvature\footnote{In eq. (19) of the second paper of ref.
\cite{Howe} the factor of 1/3 should be replaced by 1/24. We thank P. Howe for
a communication concerning this.}:
\bea
\label{torsolutions}
(e)~~~&&{T_{a \b}} ^\gamma= \frac{1}{36}{(\d_a^b\Gamma^{cde}+\frac{1}{8}{\Gamma_a}^{bcde})_{\b}}^\gamma H_{bcde}
\nonumber\\
%(e)~~~&&T_{a \b \gamma}= \frac{1}{36}H_{abcd}(\Gamma^{bcd})_{\b \gamma}
%+\frac{1}{288}(\Gamma_{abcde})_{\b \gamma}H^{bcde}
%\nonumber\\
(f)~~~&&{T_{ab}}^\a=\frac{i}{42}{(\Gamma^{cd})}^{\a\b} \Del_\b H_{abcd}\\
(g)~~~&&(\Gamma^{abc})_{\a\b}{T_{bc}}^ \b=0\nonumber
\eea
\bea
\label{cursolutions}
(h)~~~&&{R_{ab,\gamma}} ^\d= \Del_a{T_{b \gamma}}^\d-\Del_b{T_{a\gamma}}^\d +\Del_{\gamma}
{T_{ab}}^\d+{T_{a \gamma}}^\e {T_{b \e}}^\d-{T_{b \gamma}}^\e {T_{a \e}}^ \d
\nonumber\\
(i)~~~&&R_{\a b,cd}=\frac{i}{2}[(\Gamma_{b})_{\a\b}{ T_{cd }}^\b -(\Gamma_c)_{\a\b}
 {T_{db}}^\b
+(\Gamma_d)_{\a\b}{ T_{cb}}^\b ]\\
(j)~~~&&R_{\a\b,ab}= -\frac{i}{6}\left[{(\Gamma^{cd})}_{\a \b} H_{abcd} + \frac{i}{24}
(\Gamma_{abcdef})_{\a \b}H^{cdef}\right] \nonumber
\eea
with
\beq
{R_{AB \g}}^{\d} = \frac{1}{4}{R_{AB cd}}
{(\Gamma^{cd})_\g}^\d~~~.
\eeq
 
We will need the following additional consequence of the Bianchi identities:
\bea
\label{addrel}
(k)~~~&&\Del_\a H_{bcde}=- 6i(\Gamma_{[bc})_{\a\b} {T_{de]}}^\b \\
(l)~~~&&\Del_\a R_{bc,de}=\Del_bR_{\a c,de}-\Del_cR_{\a b,de}
+{T_{b\a}}^{\gamma}R_{\gamma c,de}-{T_{c\a}}^{\gamma}R_{\gamma b,de} 
-{T_{bc}}^{\beta}R_{\b \a ,de} \nonumber
\eea
which can be used to relate higher components (in $\theta$) of field strengths 
and curvatures to lower components. 

We also note the three-form equation of motion (a consequence of the
constraints)
\beq
\Del^a H_{abcd} = - \frac{1}{1728}\varepsilon_{bcd e_1 \cdots e_8}
H^{e_1 \cdots e_4} H^{e_5 \cdots e_8}
\eeq

\sect{The Supermembrane}
~~~~The supermembrane action is written in terms of superspace embedding
coordinates $Z^M(\zeta) =( x^m(\zeta),\theta^{\mu}(\zeta))$, functions of the
world-volume coordinates $\zeta^i$ ($i=0,1,2$):
\beq
S(Z)= \int d^3 \zeta [ - \sqrt{-\det G(Z) }-\frac{1}{6} \varepsilon^{ijk}\Pi_i^A\Pi_j^B\Pi_k^C
B_{CBA}(Z)]
\eeq
with $\Pi_i^A = (\pa Z^M/\pa \zeta^i) {E_M}^A(Z)$ and $G_{ij}= \Pi_i^a \Pi_j^b \eta_{ab}$.
It is $\kappa$ invariant when the background satisfies the supergravity constraints.

For the applications we have in mind, one wants to expand the action with respect to
the fermionic coordinates $\theta$ and identify the coefficients  with various
supergravity component fields: the component vielbein, the gravitino, etc. Aside
from some technical complications, this  expansion is not difficult for tensors such
as the three-form, but is highly nontrivial for the vielbein if one proceeds in
a brute force manner. As we have indicated in the introduction, previous workers
have used gauge completion to achieve this task, but it is fairly clear that this
procedure becomes quickly unmanageable if one tries to proceed   beyond
second order on the way to the full expansion at 32nd order.

Our approach, which is based on applying standard normal coordinate expansions
to this case, is systematic, recursive, covariant, and in principle could be
carried out all the way to the 32nd order - not that we recommend it. In this
approach the fermionic variables are normal coordinates on the curved 
superspace manifold
in directions ``perpendicular'' to the bosonic base manifold, and the expansion
automatically produces coefficients which are covariant objects, torsions, 
curvatures and field strengths. In general, the normal coordinates $y^\a$
are related by field redefinitions to the  ordinary $\theta^\a$ used in the gauge
completion procedure, but are equally suitable for all applications.

\sect{Normal Coordinate Expansions}
~~~~We shall rely primarily on ref. \cite{norcor1} (see also ref. \cite{sigmamodel}) and
we refer the reader to that reference for an explanation of the procedure and
some of the details. Basically, one chooses a point $Z^M=(x^m, \theta^\mu =0)$ on the
superspace manifold  as the
origin of normal coordinates, and  parametrizes  its neighborhood by coordinates
along the tangent hyperplane, $y^A= (y^a, y^\a )$. Subsequently, by setting $y^a=0$
one thus parametrizes the manifold by $(x^m, y^\a )$. One introduces an operation $\d$ whose
repeated iteration gives the successive terms in the  Taylor expansion
\beq
E^A(z;y) \equiv dZ^M{E_M}^A=E^A(Z) +\d E^A + \frac{1}{2!} \d^2 E^A +\frac{1}{3!}\d^3
E^A+...... 
\eeq
This operator acts as follows:
\beq
 \d E^A = Dy^A +y^C E^B T_{BC}^{~~~A}   \label{varviel1}
\eeq
where the covariant differential is
\beq
Dy^A \equiv  E^B \Del_B y^A = dy^A-y^B \o_B^{~A} = dy^A -y^B E^C
\o_{CB}^{~~~A} \label{eq:Dy} 
\eeq
Also
\beq
\d Dy^A = -y^B E^C y^D R_{DCB}{}^A   \label{eq:varDy}
\eeq
and for any tensor
\beq
 \d {\cal T} = y^A \Del_A {\cal T} \label{eq:vartensor}
\eeq

We record here the first few terms in the expansion of the vielbein:
\bea
\d^2E^A &=& -y^B E^Cy^D{R_{DCB}}^{A} + y^C E^B y^D \Del_D T_{BC}^{~~~A}
              \nonumber\\
       &&+ y^C(Dy^B +y^E E^D T_{DE}^{~~~B})T_{BC}^{~~~A} \label{varviel2}
\eea
\bea
\d^3 E^A &=& -y^D (Dy^B+ y^F E^G {T_{GF}}^B) y^C {R_{CBD}}^A
      -y^D E^By^C y^F \Del_F  {R_{CBD}}^A
              \nonumber\\
&&+ 2y^C (Dy^B + y^F E^G {T_{GF}}^B)y^D \Del_D {T_{BC}}^A +
         y^C E^B y^D y^E \Del_E \Del_D {T_{BC}}^A
        \nonumber\\
&& +y^C y^G(Dy^D + y^E E^F {T_{FE}}^D)
           {T_{DG}}^B {T_{BC}}^A+y^C Dy^B y^D \Del_D {T_{BC}}^A  \nonumber
\\
&&   - y^C y^D E^F y^E {R_{EFD}}^B {T_{BC}}^A
+y^C y^F E^D y^E (\Del_E {T_{DF}}^B){T_{BC}}^A
\label{varviel3}
\eea
The next term is:
\bea
\label{varviel4}
\d^4 E^A &=&  +3 y^C y^F(Dy^E + y^G E^H {T_{HG}}^E){T_{EF}}^B y^D \Del_D {T_{BC}}^A \non\\
&&  +3y^C(Dy^B + y^F E^G {T_{GF}}^B) y^D y^E \Del_E \Del_D {T_{BC}}^A  \non \\
&& - y^C y^D (Dy^F + y^G E^H {T_{HG}}^F) y^E {R_{EFD}}^B  {T_{BC}}^A \non \\
&& +2 y^C y^F (Dy^D +y^EE^G{T_{GE}}^D)y^H (\Del_H {T_{DF}}^B) {T_{BC}}^A   \non \\
&& - y^D y^F(Dy^E + y^H E^G {T_{GH}}^E){T_{EF}}^B y^C {R_{CBD}}^A \non \\
 &&  -2 y^D (Dy^B + y^F E^G {T_{GF}}^B) y^C y^E \Del_E {R_{CBD}}^A  \non \\
&& + y^C y^G y^E ( Dy^F + y^I E^H {T_{HI}}^F)  {T_{FE}}^D {T_{DG}}^B {T_{BC}}^A \\
&&+ y^D y^E E^F y^G {R_{GFE}}^B y^C {R_{CBD}}^A
- y^D y^F E^Gy^E(\Del_E{T_{GF}}^B) y^C {R_{CBD}}^A\non\\
&& - y^D E^B y^C y^E y^F \Del_F \Del_E {R_{CBD}}^A 
 + y^C E^B y^D y^E y^F \Del_F \Del_E \Del_D {T_{BC}}^A \non \\
&& -y^C y^D E^F y^E y^G (\Del_G  {R_{EFD}}^B )  {T_{BC}}^A
 -3y^C y^D E^F y^E {R_{EFD}}^B y^G \Del_G {T_{BC}}^A \non \\
&&  - y^C y^E y^F E^G y^H {R_{HGF}}^D {T_{DE}}^B {T_{BC}}^A 
 +  y^C y^G y^E E^F y^H (\Del_H {T_{FE}}^D) {T_{DG}}^B {T_{BC}}^A\non \\
&& + y^C y^F E^D y^E y^G(\Del_G \Del_E {T_{DF}}^B ){T_{BC}}^A 
 +3 y^C y^F E^D y^E (\Del_E {T_{DF}}^B) y^G \Del_G {T_{BC}}^A  \non 
\eea
Higher order
terms can be obtained by applying the $\d$ operation iteratively.
Lest the reader be frightened by the large number of terms one generates,
let  us point out that the supergravity constraints render many of the terms in the
expansion zero, as we shall see in the next section.

The pullbacks $\Pi_i^A = \pa_i Z^M{E_M}^A$ have an equivalent expansion\footnote
{With our conventions, there is a sign difference from these references.}
\cite{sigmamodel,bfunction}, with
\beq
\d \Pi_i^A =D_i y^A +y^B \Pi_i^C{T_{CB}}^A
\eeq
and
\bea
D_iy^A&=& \pa_iy^A -y^C \Pi_i^B {\omega_{BC}}^A \nonumber\\
\d D_iy^A &=& -y^B \Pi_i^C y^D R_{DCB}{}^A   \label{eq:varDy}
\eea
so that the successive terms can be read immediately from  (\ref{varviel1},
\ref{varviel2}- \ref{varviel4}).

In the above formulas  $ {\omega_{BC}}^A$ is the supergravity Lorentz connection;
this noncovariant term will drop out in any expression which is Lorentz invariant.
Furthermore, it would appear that in the expansion of the three-form, according to (4.5) 
non-gauge-invariant  supergravity covariant derivatives $\Del_A B_{BCD}$ would appear.
However, it is not difficult to  check that in the expansion of the Wess-Zumino term
in the action, from the  first variation of the vielbein factors, one generates
torsion factors so that in fact
\beq
\d[ \varepsilon^{ijk}\Pi_i^A\Pi_j^B\Pi_k^C
B_{CBA}(Z)] = \varepsilon^{ijk}y^A \Pi_i^B\Pi_j^C\Pi_k^D H_{DCBA}
\label{deltawz}
\eeq
up to total derivatives.
Subsequent terms in the expansion involve only the gauge-invariant
field strength.

For the actual application we have in mind, the expansion of the
supermembrane action in powers of the fermionic variables, the
following simplifications occur, in addition to those due to the
supergravity constraints: 
\begin{itemize}
\vspace{-0.1in}\item $y^A = (0, y^\a)$\\
\vspace{-0.2in}\item All quantities in the expansion are evaluated at $Z^M= (x^m,0)$ so that they
just involve the $\theta=0$, first components, of the superfields and their
derivatives. In particular we have (in Wess-Zumino gauge but in fact
our expansion is completely supergravity gauge-covariant), with ${E_M}^A| = 
{E_M}^A(x,0)$
\bea
{E_m}^a|&=&{e_m}^a (x) \nonumber\\
{E_m}^\a |&=& \psi_m^\a(x) \nonumber\\
{E_\mu}^a|&=&0 \nonumber\\
{E_\mu}^\a |&=& \d_\mu^\a
\eea
as well as
\bea
{T_{cd}}^\a | = \hat{\psi}_{cd}^\a
\eea
the {\em supercovariantized} gravitino field strength. We also note that
the choice of Wess-Zumino gauge  implies that the  spinor covariant
derivative connection vanishes at $\theta=0$, (see, for example, {\em Superspace}, 
 eq. (5.6.8) \cite{Superspace}.)
\beq
{\omega_{\a\b}}^\g | =0~~~~~~~~
\eeq
\vspace{-0.1in}\item One is often interested in bosonic backgrounds, and in
that case any individual
object containing an odd number of fermionic indices can be set to zero.
\end{itemize}

\sect{The  Expansion of the Vielbein}
~~~~We illustrate the remarks at the end of the previous section by
examining the first few terms in the expansion. To begin with, we
consider the bosonic  $E^a$ and take advantage of the constraints.
We find, for the lowest orders
\bea
\d E^a&=& y^\b E^C {T_{C\b}}^a =-iy^\b E^{\gamma}(\Gamma^a)_{\b \g}
=-id x^m (y \Gamma^a\psi_m)\\
&~&\non\\
\d^2 E^a&=&-y^\b E^Cy^\d {R_{\d C \b}}^a 
+ y^\g E^By^\d \Del_\d {T_{B\g}}^a
+y^\g(D y^\b \d_\b^B +y^\e E^D {T_{D\e}}^B){T_{B\g}}^a \nonumber\\
&=&-iy^\g D y^\b (\Gamma^a)_{\b \g}
 -iy^\g E^d y^\e{T_{d\e}}^\b(\Gamma^a)_{\b \g}\nonumber\\
&=&-i (y \Gamma^a D y) - idx^m{e_m}^d  y^\g y^\e{T_{d\e}}^\b (\G^a)_{\b \g}\\
&~&\non\\
\d^3 E^a&=& iy^\g y^\d  dx^m\psi_m^\phi y^\e {R_{\e \phi \d}}^\b(\G^a)_{\b \g}
+iy^\g y^\d dx^m{e_m}^f y^\e {R_{\e f \d}}^\b(\G^a)_{\b \g} \\
&&- y^\g y^\kappa y^\e d x^m \psi_m^\phi (\G^d)_{\phi \e} {T_{d \kappa}}^\b (\G^a)_{\b \g} 
 -i y^\g y^\phi dx^m{e_m}^d y^\e (\Del_\e {T_{d \phi}}^\b ) (\G^a)_{\b \g}\nonumber
\eea

In a similar manner, for the fermionic $E^\a$,
\bea
\d E^\a &=& Dy^\a +y^\g dx^m{e_m}^b{T_{b\g}}^\a\\
&~& \non\\
\d^2 E^\a &=& -y^\b dx^m {e_m}^c y^\d {R_{\d c \b}}^\a -y^\b dx^m \psi_m^\g y^\d {R_{\d \g \b}}^\a
\nonumber\\
&&+y^\g dx^m {e_m}^b y^\d \Del_\d {T_{b\g}}^\a -iy^\g  dx^m (y \G^b \psi^m){T_{b\g}}^\a\\
&~&\non\\
\d^3 E^\a &=&-y^\d Dy^\b y^\g {R_{\g \b \d}}^\a - y^\d y^\phi dx^m({e_m}^c {T_{c \phi}}^\b y^\g {R_{\g \b \d}}^\a
 -i\psi_m^\kappa (\Gamma^b)_{\kappa \phi}y^\g {R_{\g b \d}}^\a )\nonumber\\
&&-y^\d dx^m(\psi_m^\b y^\g y^\phi \Del_\phi {R_{\g \b \d}}^\a +{e_m}^b y^\g y^\phi \Del_\phi {R_{\g b \d}}^\a )\non\\
&& -i y^\g y^\phi dx^m \psi_m^\kappa (\Gamma^b)_{\kappa \phi} y^\d \Del_\d {T_{b\g}}^\a \\
&&+y^\g dx^m {e_m}^b y^\d y^\e \Del_\e \Del_\d {T_{b\g}}^\a
-iy^\g y^\kappa Dy^\d (\Gamma^b)_{\d \kappa}{T_{b\g}}^\a  \non\\
&&-iy^\g y^\kappa y^\e dx^m {e_m}^f {T_{f\e}}^\d(\Gamma^b)_{\d \kappa}{T_{b \g}}^\a 
-iy^\g y^\phi dx^m \psi_m^\d (\Gamma^b)_{\d \phi}y^\e \Del_\e {T_{b \g}}^\a \nonumber
\eea
etc. All the quantities on the right-hand side are evaluated at $\theta =0$.

We have used the constraints, and also the structure of the tangent space Lorentz
group which implies ${R_{AB \g}}^a=0$. The torsion $ {T_{d \a}}^\b$
 is given by the constraints (\ref{torsolutions})
in terms of the  three-form field strength  {\it evaluated at} $\theta=0$, i.e. the
component  field strength. We note  the vanishing of  $\Del_\d {T_{\b \g}}^a $.
   We also note the appearance of the component vielbein
and of the gravitino. As we show in the next section the other quantities, 
all evaluated at  $\theta =0$, can be
 expressed, via the solution of the Bianchi identities, in terms of component
quantities.

The expansion of the individual vielbein components, ${E_M}^A$ can be read
from the  above expansion of $E^A = {E_m}^A dx^m + {E_\mu}^A dy^\mu$ 
(in WZ gauge $y^\a = \d_\m^\a y^\mu$). Thus, to read off the expansion
of  ${E_\mu}^A$ at a given
order in $y$, one has to go one order higher for  $E^A$.  Although in our
approach these components are not needed for the expansion of the
membrane, we give the first few orders.
We obtain, from (5.1-3) (and the $Dy$ part of  (\ref{varviel4}))
\bea
{E_m}^a &=& {e_m}^a (x)  -i (y \G^a\psi_m)\nonumber\\
&&-\frac{i}{2}\left[{ y^\g y^\e e_m}^d {T_{d\e}}^\b
(\G^a)_{\b \g} + y^\a (\G^a)_{\a\b} y^\g {e_m}^c {\o_{c \g}}^\b \right] \nonumber\\
&&+\frac{1}{3!} \left[ iy^\g y^\d \psi_m^\phi y^\e {R_{\e \phi \d}}^\b(\G^a)_{\b \g}
+iy^\g y^\d{e_m}^f y^\e {R_{\e f \d}}^\b (\G^a)_{\b \g}  \right. \\
 &&\left.- y^\g y^\kappa y^\e \psi_m^\phi (\G^d)_{\phi \e} {T_{d \kappa}}^\b (\G^a)_{\b \g} 
 -i y^\g y^\phi {e_m}^d y^\e (\Del_\e {T_{d \phi}}^\b ) (\G^a)_{\b \g} \right] +O(y^4)
\nonumber\\
&~& \non\\
{E_\m}^a &=& -\frac{i}{2} y^\g(\G^a)_{\g\b} \d_\m^\b  \\
&&+\frac{1}{4!}\d_\mu^\b \left[y^\g y^\kappa y^\e (\G^d)_{\b \e} {T_{d \kappa}}^\phi
(\G^a)_{\phi \g} + i y^\g y^\d y^\e {R_{\e \b \d}}^\phi (\G^a)_{\phi \g} \right]  +O(y^5)\nonumber
\eea
and from (5.4-6)
\bea
{E_m}^\a &=& \psi_m^\a - iy^\b {e_m}^c {\o_{c \b}}^\a +y^\g {e_m}^b {T_{b\g}}^\a
\nonumber\\
&&+\frac{1}{2}\left[ -y^\b  {e_m}^c y^\d {R_{\d c \b}}^\a -y^\b  \psi_m^\g y^\d {R_{\d \g \b}}^\a
+y^\g  {e_m}^b y^\d\Del_\d  {T_{b\g}}^\a -iy^\g   (y \G^b \psi_m){T_{b\g}}^\a \right] \nonumber\\
&&+\frac{1}{3!} \left[y^\d y^\e {e_m}^c{\o_{c\e}}^\b y^\g{R_{\g\b\d}}^\a  - y^\d y^\phi ({e_m}^c {T_{c \phi}}^\b y^\g {R_{\g \b \d}}^\a
 -i\psi_m^\kappa (\Gamma^b)_{\kappa \phi}y^\g {R_{\g b \d}}^\a )\right.\nonumber\\
&&\left.-y^\d (\psi_m^\b y^\g y^\phi \Del_\phi {R_{\g \b \d}}^\a +{e_m}^b y^\g y^\phi \Del_\phi {R_{\g b \d}}^\a )
 -i y^\g y^\phi  \psi_m^\kappa (\Gamma^b)_{\kappa \phi} y^\d \Del_\d {T_{b\g}}^\a\right. \non \\
&&\left.+y^\g  {e_m}^b y^\d y^\e \Del_\e \Del_\d {T_{b\g}}^\a
   +iy^\g y^\phi y^\b {e_m}^c {\o_{c\b}}^\d (\G^b)_{\d \phi} {T_{b \g}}^\a                     \right.  \\
&&\left.-iy^\g y^\kappa y^\e {e_m}^f {T_{f\e}}^\d(\Gamma^b)_{\d \kappa}{T_{b \g}}^\a-iy^\g y^\phi  \psi_m^\d (\Gamma^b)_{\d \phi}y^\e \Del_\e {T_{b \g}}^\a \right] +O(y^4)  \non\\
&~&\non\\
{E_\m}^\a &=& \d_\m^\a +\frac{1}{2} \left[ \d_\m^\b y^\d y^\g {R_{\g\b\d}}^\a
-i\d_\m^\d y^\g y^\kappa (\G^b)_{\d \kappa}{T_{b \g}}^\a \right]  +O(y^4)
\eea
We describe in the next section how the various superspace tensors appearing above, which are evaluated at $\theta=0$, are to be identified with component fields.

The first two orders of the
expansion  can be compared with,
for example,  (4.15, 5.1) of ref. \cite{Bernard}, although the authors of that reference,
using gauge completion,  were not able to obtain all the ${\cal O}(\theta^2)$ terms.

\sect{The Component Fields}
~~~~In the general expansion we encounter  superspace fields and their
spinor derivatives evaluated at $\theta =0$ and we must express these
quantities in terms of the fields of component supergravity.

{}From the  relations in (\ref{torsolutions}, \ref{cursolutions})  it is clear that 
all  torsions and  two of the curvatures
can be expressed in terms of the field strength $H_{abcd}(x, \theta)$.
We identify
\beq
H_{abcd}| = \hat{h}_{abcd}(x)
\eeq
where $\hat{h}_{abcd}$ is the {\em supercovariantized} component field strength.
Similarly, for the curvature\beq
{R_{ab \g}}^\d| = {\hat{r}_{ab \g}}^\d(x)
\eeq
the  supercovariant component curvature tensor. 

We will need the $\theta=0$ spinor derivatives of some of these quantities.
 To begin with, from (l) in eq. (\ref{addrel})
\beq
\Del_\a H_{abcd}|
 = 6i{(\Gamma_{[ab})_\a}^\b T_{cd]\b} |=6i{(\Gamma_{[ab})_\a}^\b \hat{\psi}_{cd]\b}(x) 
\eeq
in terms of the  supercovariantized gravitino field strength. 
Taking a second spinor derivative,
\beq
\Del_\g\Del_\a H_{abcd}|
 = 6i{(\Gamma_{[ab})_\a}^\b \Del_\g T_{cd]\b} |
\eeq
and  at $\theta=0$ the spinor derivative of the  torsion can be expressed, by means of
(h) in (\ref{cursolutions}) in terms of the component curvature, $\Del_a{T_{b \g}}^ \d \sim \Del_a H_{bcde}$ ($\Del_a H_{bcde}| = \hat{D}_a \hat{h}_{bcde}$ is
a supercovariantized space-time derivative), and the products of two $H$. Clearly this
procedure can be repeated so that at every stage spinor derivatives of
 $H_{abcd}$ can be expressed in terms of the component curvature, gravitino field strength,
$\hat{h}_{abcd}$ , and space-time derivatives thereof.

Turning to the curvatures, a similar procedure applies. Relations (i) and (j) in
(\ref{cursolutions}) allow us to express curvatures with some spinor indices in terms
of the component quantities above. The relation (l) in (\ref{addrel}) gives us
the first spinor derivative of the curvature with vector indices.  A second spinor
derivative will lead to terms such as $\Del_\g \Del_b R_{\a c ,de}| =
\Del_b \Del_\g R_{\a c ,de}| +[\Del_\g ,\Del_b ]R_{\a c ,de}|$ and the
commutator can  be expressed  in terms of  known torsions and curvatures.

The procedure we have outlined above provides a systematic and unambiguous
way of obtaining the component expansions to any order. Obviously higher orders
will get more and more complicated, but they don't require any ingenuity, just
straightforward application of the rules we have described.

\sect{Expansions in bosonic backgrounds}
~~~~The higher order expansions simplify considerably if we consider
purely bosonic backgrounds - the case often of  interest. 
The main simplification comes about because we can set to zero all
quantities with an odd number of spinor indices. In particular, it is
evident that odd terms in the expansion of $E^a$ and even terms in the
expansion of $E^\a$ vanish. It is then a simple matter to examine the general
expansion  and write down
the actual form for this case. Thus we find, from the general expansion in Section 4
\bea
\d E^a &=& 0\\
&~& \non\\
\d^2 E^a &=& -i(y \G^a Dy) -i y^\g y^\e e^d {T_{d \e}}^\b {(\G^a)_{\b \g}}\\
&~&\non\\
\d^3 E^a &=&0 \\
&~&\non\\
\d^4 E^a &=& iy^\g y^\d e^f y^\e y^\kappa (\Del_\kappa {R_{\e f \d}}^\b)(\G^a)_{\b \g}
- y^\g y^\kappa y^\e Dy^\phi( \G^d)_{\phi \e} {T_{d \kappa}}^\b (\G^a)_{\b \g}\\
&&-y^\g y^\nu y^\e y^\l e^h {T_{h \l}}^\phi (\G^d)_{\phi \e} {T_{d \nu}}^\b (\G^a)_{\b \g}
-i y^\g y^\d e^f y^\e y^\nu (\Del_\nu \Del_\d {T_{f\phi}}^\b ) (\G^a)_{ \b \g} \nonumber
\label{bosexp}
\eea
and
\bea
\d E^\a &=& Dy^\a +y^\g e^b {T_{b \g}}^\a\\
&~&\non\\
\d^2E^\a &=&0\\
&~&\non\\
\d^3E^\a&=&-y^\d Dy^\b y^\g[ {R_{\g\b\d}}^\a +i (\G^b)_{\b\d}{T_{b\g}}^\a] \\
&&-y^\d y^\phi y^\g e^b[{T_{b\phi}}^\b {R_{\g\b\d}}^\a -\Del_\g{R_{\phi b\d}}^\a
-\Del_\phi\Del_\d {T_{b\g}}^\a +i {T_{b\gamma}}^\kappa (\G^b)_{\kappa \phi} {T_{b\d}}^\a]\nonumber
\eea
In these expressions several quantities appear (evaluated at $\theta=0$)
which must be worked out using the Bianchi identities and their 
consequences, as explained earlier. We proceed in the following manner:

The torsion $ {T_{a\b}}^\g$ is expressed directly, from (\ref{torsolutions}) in terms of
the field strength $H_{abcd}$ which at $\theta=0$ is the component field
strength.

The spinor derivative of the spinor-vector curvature is obtained as follows:
\bea
\Del_\phi {R_{\g b \d}}^\a &=& \frac{1}{4} \Del_\phi R_{\g bcd}{(\G^{cd})_\d}^\a
\nonumber\\
&=&\frac{i}{8} {(\G^{cd})_{\d}}^\a[(\G_b)_{\g\b} \Del_\phi {T_{cd}}^\b 
-(\G_c)_{\g\b} \Del_\phi {T_{db}}^\b+(\G_d)_{\g\b} \Del_\phi{T_{cb}}^\b]
\nonumber\\
&=&\frac{i}{8}{(\G^{cd})_{\d}}^{\a}[(\G_b)_{\g\b} {R_{cd\phi}}^\b -2 \Del_c {T_{d\phi}}^\b
-2{T_{c\phi}}^\e {T_{d \e}}^\b ) \\
&&~~~~-2(\G_c)_{\g\b}(
{R_{db\phi}}^\b- \Del_d {T_{b\phi}}^\b+\Del_b {T_{d\phi}}^\b
-{T_{d\phi}}^\e {T_{b \e}}^\b +{T_{b\phi}}^\e {T_{d \e}}^\b)]\nonumber
\eea 
where we have used  the relations (\ref{addrel})  and the solution (h) to the Bianchi
identities to determine objects such as $ \Del_\phi T_{cd\b} $. In the final expression
all quantities, evaluated at $\theta=0$ involve the component curvature, the
component  four-form field strength and space-derivatives thereof.

In a similar fashion we have
\bea
\Del_\e \Del_\d {T_{b\g}}^\a &=& \frac{1}{36}(\d_b^a\Gamma^{cde}+\frac{1}{8}{\Gamma_b}^{acde}{)_\g}^\a
\Del_e \Del_\d H_{acde} \nonumber\\
&=&-\frac{i}{6}(\d_b^a\Gamma^{cde}+\frac{1}{8}{\Gamma_b}^{acde}{)_\g}^\a
(\G_{[ac})_{\d\b} \Del_\e {T_{de]}}^\b    \\
&=&-\frac{i}{6}(\d_b^a\Gamma^{cde}+\frac{1}{8}{\Gamma_b}^{acde}{)_{\g}} ^\a
\G_{[ac})_{\d\b} \nonumber\\
&& \cdot
[{R_{de]\e}}^\b-\Del_d{T_{e]\e}}^\b+\Del_e {T_{d] \e}}^\b- {T_{d|\e|}}^\g {T_{e]\g}}^\b +{T_{e|\e|}}^\g {T_{d]\g}}^\b]
\nonumber
\eea
and again,  at $\theta=0$, the quantities in the last line are recognizable component curvatures,
field strengths, and space-derivatives thereof.

\sect{The expansion of the brane action}

The first ingredient one needs is the expansion of the first
term in the membrane action, $S_\s= \sqrt{-\det G_{ij}} =\sqrt{ - \det(\Pi_i^a \Pi_j^a)}$.
The general expansion is straightforward though lengthy. We quote the first
few terms
\bea
\d S_\s &=& \frac{1}{2}S_\s (\tr G^{-1}\d G)\\
&~& \non\\
\d^2S_\s &=&\frac{1}{4}S_\s (\tr G^{-1}\d G)^2 -\frac{1}{2}S_\s(\tr G^{-1} \d G G^{-1} \d G 
-\tr G^{-1}\d^2G)\\
&~&\non\\\d^3 S_\s &=&\frac{1}{8}S_\s (\tr G^{-1}\d G)^3 +\frac{3}{4} S_\s(\tr G^{-1} \d G)(\tr G^{-1}\d^2G)
\nonumber\\
&&-\frac{3}{4}S_\s(\tr G^{-1} \d G)(\tr G^{-1} \d G G^{-1} \d G) +S_\s(\tr G^{-1}\d GG^{-1}\d G
G^{-1} \d G) \nonumber\\
&&-\frac{3}{2}S_\s (\tr G^{-1} \d^2GG^{-1}\d G) +\frac{1}{2}S_\s (\tr G^{-1} \d^3 G)
\eea
with $\d G_{ij}= \d \Pi_i^a \Pi_j^a +\Pi_i^a \d \Pi_j^a$.
The next term in the expansion can be obtained straightforwardly.

Next we consider the expansion of the Wess-Zumino term. Starting with the
first order  variation (\ref{deltawz}) one applies the $\d$ operation repeatedly.
Thus one finds, evaluating at $\theta =0$, with
\bea
\Pi_i^a| &=& \pa_ix^m {e_m}^a(x) = \pi_i^a\\
G_{ij}| &=&\Pi_i^a \Pi_j^a| =\pi_i^a \pi_j^a \equiv g_{ij}(x) \nonumber
\eea
the following:
\bea
\d(WZ)&=& \frac{i}{2}\e^{ijk}y^\a \pa_ix^m \pi_j^c\pi_k^d \psi_m^\b (\G_{dc})_{\b \a} \\
&~&\non\\
\d^2(WZ)&=& \frac{1}{6}\e^{ijk}y^\a [3 (\d \Pi_i^B)\Pi_j^C\Pi_k^D H_{DCB\a}
+\Pi_i^B\Pi_j^C\Pi_k^D y^\l \Del_\l H_{DCB\a}] |\\
&=&\frac{1}{2}i\e^{ijk}y^\a [(\d \Pi_i^\b|) \pi_j^c \pi_k^d (\G_{dc})_{\b \a} +2 (\d\Pi_i^b|)\pi_j^c \pa_kx^m\psi_m^\d (\G_{cb})_{\d \a}] \nonumber\\
&~& \non\\
\d^3(WZ)&=&\frac{1}{6} \e^{ijk}y^\a [3(\d^2 \Pi_i^B)\Pi_j^C\Pi_k^D H_{DCB\a}
+6(\d \Pi_i^B)(\d \Pi_j^C)\Pi_k^D   H_{DCB\a}\nonumber \\
&&+6(\d \Pi_i^B)\Pi_j^C\Pi_k^D y^\l \Del_\l H_{DCB\a}
+\Pi_i^B\Pi_j^C\Pi_k^D y^\m y^\l \Del_\l \Del_\m H_{DCB\a}]|\nonumber\\
&=&\frac{1}{2}i\e^{ijk}y^\a[(\d^2 \Pi_i^\b|) \pi_j^c\pi_k^d (\G_{dc})_{\b \a}
+2(\d^2\Pi_i^b|) \pi_j^c\pa_kx^m\psi_m^\d (\G_{cb})_{\d \a} \nonumber\\
&&+4 (\d \Pi_i^\b|)(\d\Pi_j^c|) \pi_k^d (\G_{dc})_{\b \a})
+2(\d\Pi_i^b|)( \d\Pi_j^c|) \pa_kx^m \psi_m^\d (\G_{cb})_{\d \a}]
\eea
etc. The $\d^n \Pi_i^A$ can be read from the $\d^n E^A$ in Section 5.

\subsection{The membrane up to second order}

It is now a straightforward matter to determine the low-order expansion of the brane action.
 We have
\bea
S^{(0)}(x;y)&=& \int d^3 \zeta \left[ - \sqrt{-g(x) }-\frac{1}{6} \varepsilon^{ijk}\pi_i^a\pi_j^b\pi_k^c
b_{cba}(x)\right]  \\
&~& \non\\
S^{(1)}(x;y)&=& \int d^3 \zeta\left[-i\sqrt{-g}g^{ij}\pi_i^a \pa_jx^n (y \Gamma_a \psi_n)
+\frac{1}{2}i \e^{ijk} \pi_i^a \pi_j^b \pa_k x^n (y \Gamma_{ab} \psi_n)\right] \nonumber\\
&~&\\
S^{(2)}(x;y) &=&  \int d^3 \zeta \left[\frac{1}{2}\sqrt{-g} \left \{- [g^{ij}\pi_i^a\pa_jx^n(y\Gamma_a\psi_m)]^2 +
2 g^{ij}g^{kl} \pi_j^a\pi_l^b \pa_kx^m\pa_ix^n(y\Gamma_a\psi_m)(y\Gamma_b\psi_n)
\right.\right.\nonumber\\
&&\left.\left. -g^{ij}\pa_ix^m (y \G^a \psi_m)\pa_jx^n(y\G_a \psi_n)
 -ig^{ij}\pi_i^a(y\Gamma_aD_jy) -i g^{ij}\pi_i^a\pi_j^by^\g  y^\e {T_{b\e}}^\b (\Gamma_a)_{\b \g} \right\} \right.\nonumber\\
&&\left.-\frac{1}{4}i\e^{ijk}\left\{ \pi_i^a \pi_j^b(y \Gamma_{ba}D_ky) +\pi_i^a \pi_j^b \pi_k^c y^\a y^\g {T_{a\g}}^\b (\Gamma_{cb})_{\b \a} \right.\right.\nonumber\\
&&\left.\left.+2i \pi_i^a \pa_jx^m \pa_kx^n (y \G^b \psi_m)(y \G_{ab}\psi_n)\right \}\right]
\eea
In these expressions we must substitute for the torsion $ {T_{b\e}}^\b$ as given in (\ref{torsolutions}).

The result above could be used for one-loop calculations and for identifying
gravitino emission vetices \cite{Plefka}.

\subsection{The membrane in bosonic backgrounds}

Matters simplify in a bosonic background where we can take advantage of the
vanishing of odd variations of $\Pi_i^a$ which implies $\d G = \d^3G=0$. One finds then
\bea
\d S_\s &=& 0\nonumber\\
\d^2 S_\s &=& \frac{1}{2}S_\s(\tr G^{-1}\d^2G)\nonumber\\
\d^3S_\s &=&0 \\
\d^4S_\s&=& \frac{3}{4}S_\s(\tr G^{-1} \d^2G)^2 -\frac{3}{2}S_\s(\tr G^{-1} \d^2GG^{-1}\d^2G)
+\frac{1}{2}S_\s( \tr G^{-1} \d^4G)
\nonumber
\label{Sexp}
\eea
where 
\bea
\d^2 G_{ij} &=& \Pi_i^a \d^2 \Pi_j^a  + \d^2\Pi_i^a  \Pi_j^a\\
%&=& 2 \pi_i^a[-i(y \G^a D_jy) -
% i \pi_j^d  y^\g y^\e {T_{d\e}}^\b (\G^a)_{\b \g}] + (i\leftrightarrow j) \non\\
&~& \non\\
\d^4 G_{ij} &=& \Pi_i^a \d^4 \Pi_j^a+2\d^2\Pi_i^a \d^2 \Pi_j^a+\d^4\Pi_i^a  \Pi_j^a
\label{WZexp}
\eea

For the Wess-Zumino term  one finds  a very simple result to this order:
\bea
\d (WZ) &=&0 
\nonumber\\
\d^2 (WZ)&=&\frac{ i}{2}
\e^{ijk} y^\a \d\Pi_i^\b \Pi_j^c \Pi_k^d (\G_{cd})_{\a\b}
\nonumber\\
\d^3 (WZ) &=&0\\
\d^4 (WZ) &=& \frac{i}{2} \e^{ijk}y^\a [\d^3 \Pi_i^\b \Pi_j^c \Pi_k^d (\G_{cd})_{\a\b} +
\d\Pi_i^\b \d^2\Pi_j^c \Pi_k^d (\G_{cd})_{\a\b} ]\nonumber
\eea

The second order expansion of the action can be read immediately from (8.10)
by setting the gravitino to zero:
\bea
S^{(2)}(x;y) &=& \frac{1}{2} \int d^3 \zeta \left[-i\sqrt{-g} g^{ij}\left\{ \pi_i^a(y\Gamma_aD_jy) + \pi_i^a\pi_j^by^\g  y^\e {T_{b\e}}^\b (\Gamma_a)_{\b \g} \right\} \right. \nonumber\\
&&~~~~~~~~\left.-\frac{1}{2}i\e^{ijk}\left\{ \pi_i^a \pi_j^b(y \Gamma_{ba}D_ky) +\pi_i^a \pi_j^b \pi_k^c y^\a y^\g {T_{a\g}}^\b (\Gamma_{cb})_{\b \a} \right\}\right] \nonumber\\
&&{}
\eea

{}From (8.11-14) the fourth order expansion can be written as
\bea
S^{(4)}(x:y) &=&\frac{1}{4!} \int d^3 \zeta \left[ -\sqrt{-g} \left\{3 [\pi^{ia}\pi^{jb} - \pi^{ja}\pi^{ib}
-g^{ij}\pi^{ka}\pi_k^b + \frac{1}{3}g^{ij}\eta^{ab}]\d^2\Pi_i^a \d^2\Pi_j^b \right. \right.
\nonumber\\
 &&~~~~~~~~~\left.\left.+g^{ij}\pi_i^a \d^4\Pi_j^a \right\}-\frac{i}{2} \e^{ijk}y^\a [\d^3 \Pi_i^\b \pi_j^c +\d\Pi_i^\b \d^2\Pi_j^c]\pi_k^d(\G_{cd})_{\a\b}\right] \nonumber\\
\eea
In this expression the variations  $\d^n \Pi_i^A$ must be substituted.
As explained earlier, they  can be read from the corresponding
variations of the vielbein in Section 7. Explicitly,
\bea
\d^2 \Pi_i^a &=& -i(y \G^a D_iy) -i y^\g y^\e \pi_i^d {T_{d \e}}^\b {(\G^a)_{\b \g}}\non\\
\d^4 \Pi_i^a &=& iy^\g y^\d \pi_i^f y^\e y^\kappa (\Del_\kappa {R_{\e f \d}}^\b)(\G^a)_{\b \g}
- y^\g y^\kappa y^\e D_iy^\phi( \G^d)_{\phi \e} {T_{d \kappa}}^\b (\G^a)_{\b \g} \non\\
&&-y^\g y^\nu y^\e y^\l \pi_i^h {T_{h \l}}^\phi (\G^d)_{\phi \e} {T_{d \nu}}^\b (\G^a)_{\b \g}
-i y^\g y^\d \pi_i^f y^\e y^\nu (\Del_\nu \Del_\d {T_{f\phi}}^\b ) (\G^a)_{ \b \g} \nonumber\\
\d \Pi_i^\a &=& D_iy^\a +y^\g \pi_i^b {T_{b \g}}^\a\non\\
\d^3\Pi_i^\a&=&-y^\d D_iy^\b y^\g[ {R_{\g\b\d}}^\a +i (\G^b)_{\b\d}{T_{b\g}}^\a] \\
&&-y^\d y^\phi y^\g \pi_i^b[{T_{b\phi}}^\b {R_{\g\b\d}}^\a -\Del_\g{R_{\phi b\d}}^\a
-\Del_\phi\Del_\d {T_{b\g}}^\a +i {T_{b\gamma}}^\kappa (\G^b)_{\kappa \phi} {T_{b\d}}^\a]\nonumber
\eea
The torsions and curvatures that appear in these expressions, all evaluated
at $\theta=0$, are given in eqs. (2.4, 7.8, 7.9) in terms of component curvatures,
field strengths and space-derivatives thereof. 
We will refrain from  carrying out the substitution, or giving the
next order in the expansion

\newpage

\sect{Conclusions }

We have presented the ingredients for expanding the 11-dimensional membrane
action in powers of the superspace fermionic coordinates. In a future publication
we intend to carry out similar work for various branes in 10-dimensional
supergravity backgrounds. Our main tool has been the superspace normal
coordinate expansion suitably adapted to our application. This expansion
allows for a systematic and unambiguous method for proceeding to any order,
using as its main ingredient the geometrical formulation of the background superspace,
i.e. the constraints on torsions, curvatures, and 3-form field strength. We have used 
the standard constraints of refs. \cite{Howe, Cremmer} but the procedure is
flexible enough to allow for modifications such as those considered  by
Cederwall {\em et al} \cite{Cederwall} (see also \cite{Peeters}).

The membrane action is invariant under the various superspace symmetries, as
well as $\kappa$-symmetry. In refs. \cite{sigmamodel, bfunction}, in the context of a
background field expansion for the Green-Schwarz superstring, we have discussed how these symmetries are
reflected in symmetries of the action after the normal coordinate expansion. What we wish to
emphasize is that no additional checking of $\kappa$-invariance is necessary

Our results could be specialized to specific AdS$\times$$S$  backgrounds where the  various curvatures and field strengths become very simple. This would
allow comparison with the complete expansions obtained for such backgrounds in
refs.   \cite{Kallosh, dewit, Siegel}.  We hope to pursue this, as well as
other applications of our results, in a future publication.

{\bf Acknowledgments}\\
Part of this work was done  at the Aspen Center for Physics. We thank W. Siegel
for reading the manuscript and for many useful suggestions.
We also acknowledge conversations with B. de Wit, J. Plefka, and W. Taylor.
MTG thanks the Physics Department of  McGill University, where some of this
work was done.

\end{document}